\documentclass[showpacs,amsmath,
twocolumn,
aps,prb]{revtex4}
\usepackage{graphicx}
\usepackage{dcolumn}

\begin{document}

\title{Confinement engineering of $s$-$d$ exchange interactions in GaMnAs quantum wells}

\author{N. P. Stern}
\author{R. C. Myers}
\author{M. Poggio}
\author{A. C. Gossard}
\author{D. D. Awschalom}
\affiliation{Center for Spintronics and Quantum Computation,
University of California, Santa Barbara, CA 93106}
\date{\today}

\begin{abstract}
Recent measurements of coherent electron spin dynamics reveal an
antiferromagnetic $s$-$d$ exchange coupling between conduction band
electrons and electrons localized on Mn$^{2+}$ impurities in GaMnAs
quantum wells.  Here we discuss systematic measurements of the
$s$-$d$ exchange interaction in
Ga$_{1-x}$Mn$_{x}$As/Al$_{y}$Ga$_{1-y}$As quantum wells with
different confinement potentials using time-resolved Kerr rotation.
Extending previous investigations of the dependence of the $s$-$d$
exchange, $N_{0} \alpha$, on well width, we find that its magnitude
also depends on well depth.  Both phenomena reduce to a general
dependence on confinement energy, described by a band-mixing model
of confinement-induced kinetic exchange in the conduction band.
\end{abstract}

\maketitle

Dilute magnetic semiconductors (DMS) are a scientifically and
technologically interesting class of materials due to the strong
$sp$-$d$ exchange interactions between the $s$-like conduction band
or the $p$-like valance band and the localized $d$ shell of magnetic
dopants.\cite{Dietl:1994}  The strength of these couplings and the
resulting enhancement of Zeeman spin-splittings lead to dramatic
spin-dependent properties in DMS including the formation of magnetic
polarons \cite{Dietl:1994, Awschalom:1985}, the coherent transfer of
spin polarization from carriers to magnetic ions
\cite{Crooker:1997}, and carrier-mediated
ferromagnetism.\cite{Ohno:1996}  Heterostructures with DMS layers
offer the freedom to engineer exchange spin-splittings of carriers
in electronic devices. \cite{Myers:2005a}  The exchange interactions
in II-VI DMS have been characterized by band-edge magneto-optical
spectroscopy \cite{Twardowski:1983, Gaj:1979} and are well
understood theoretically.\cite{Larson:1988, Blinowski:1992}
Measurements in II-VI DMS quantum wells (QWs) reveal that increasing
quantum confinement reduces the strength of the $s$-$d$ exchange due
to kinetic exchange effects\cite{Mackh:1996, Bhattacharjee:1998,
Merkulov:1999}, suggesting additional avenues for manipulating
carrier exchange interactions in DMS heterostructures using
band-engineering.

Comparable studies in III-V alloys such as GaMnAs have been more
difficult because of the high defect densities in normal growth
conditions. Recent refinements of molecular-beam epitaxy (MBE)
techniques allow production of high-quality III-V paramagnetic DMS
heterostructures allowing optical measurement of exchange
interactions through electron spin coherence.\cite{Myers:2005,
Poggio:2005}  The $s$-$d$ exchange constant $N_0 \alpha$ was
observed to be antiferromagnetic in GaMnAs QWs., rather than
ferromagnetic as concluded in earlier magneto-optical studies in
bulk GaMnAs. \cite{Szczytko:1996, Ando:1998}    Despite the
difference in the sign of the interaction in GaMnAs QWs as compared
with II-VI QWs,\cite{Merkulov:1999} the data show that $N_0 \alpha$
decreases with increasing one-dimensional (1D) quantum confinement
in both materials. Here we extend previous measurements of $N_0
\alpha$ in GaMnAs/AlGaAs QWs by additionally varying the QW barrier
height, which confirms this dependence on quantum confinement.

Single Ga$_{1-x}$Mn$_{x}$As/Al$_{y}$Ga$_{1-y}$As quantum wells of
width $d$ are grown by MBE on (001) semi-insulating GaAs wafers
using the conditions outlined in Ref. \onlinecite{Poggio:2005}.  QWs
with barriers containing different fractions of Al ($y$) are grown
in which the QW barrier height $E_{\text{b}}$ is proportional to
$y$. In particular, we grow samples with $y =0.1$ and $d = 10$ nm,
with $y =0.2$ and $d = 10$ nm, and with $y =0.2$ and $d = 5$ nm.
Fig. 1a depicts the energy diagram for these structures, where the
electron kinetic energy $E_e$ is defined as the energy between the
bottom of the GaAs conduction band and the ground state energy in
the QW.  $E_e$ is calculated from the material and structural
parameters of the QWs  using a one-dimensional
Poisson-Schr\"{o}dinger solver. \cite{1D:Poisson} These structures
complement the four QWs measured for $y = 0.4$ and $d = 3$ nm, 5 nm,
7.5 nm, and 10 nm in Ref. \onlinecite{Myers:2005}. The variation of
the QW depth allows $E_e$ to be varied independently from $d$,
addressing the possibility that $d$-dependent parameters aside from
$E_e$ affect $N_0 \alpha$. By varying $y$, we circumvent possible
changes in the measured $N_0 \alpha$ due to different Mn
incorporation behavior or Mn-profile measurement artifacts related
to QW width.
\begin{figure}[b]\includegraphics{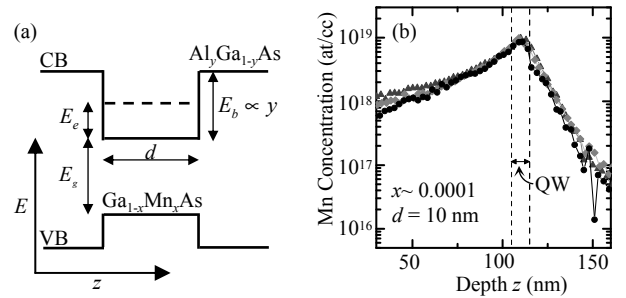}\caption{\label{fig1}
(a) Schematic of the QW band structure, showing the well width $d$,
the GaAs band gap $E_{g}$, the QW barrier height $E_b$, and the
electron kinetic energy $E_e$.  (b)  SIMS Mn profiles for three $d =
10$ nm QWs for nominally the same Mn concentration $x \sim 0.0001$,
with $y = 0.1$ (light gray), $y = 0.2$ (gray), and $y = 0.4$
(black). }\end{figure}

For each aforementioned $y$ and $d$ pair, we measure a sample set
consisting of a non-magnetic control sample ($x=0$) and four samples
with increasing Mn doping ($0.00002 <  x < 0.0007 $), for a total of
15 samples.  The effective Mn doping level $x$ is determined
quantitatively by secondary ion mass spectroscopy (SIMS) as
described in Ref \onlinecite{Poggio:2005}.  Fig. 1b shows a typical
Mn doping profile obtained by SIMS for $d=10$ nm QWs.  The
Mn-profiles in the QW region exhibit no appreciable dependence on
the Al concentration, indicating that QWs of different $y$ have
similar Mn incorporation behavior.

We observe photoluminescence from both the band-edge exciton and the
Mn acceptor (Fig. 2a).    The Mn luminescence is red-shifted by
0.107 eV from the QW band-edge, consistent with the Mn ionization
energy measured in GaAs/AlGaAs superlattices. \cite{Plot:1986}

\begin{figure}\includegraphics{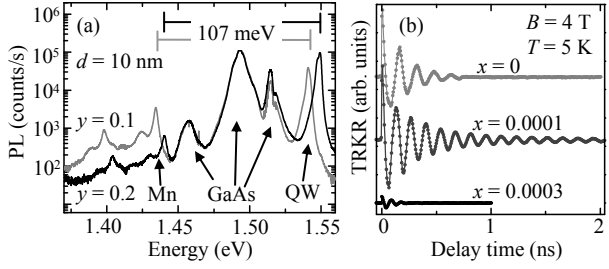}\caption{\label{fig2}
(a) PL spectra from $d = 10$ nm QWs where the Mn acceptor peak is
shifted $\sim 107$ meV lower in energy than the QW PL.  The $y =
0.1$, $x = 0.00005$ (gray) and $y = 0.2$, $x = 0.00003$ (black) QWs
are shown.  (b)  KR from $d=5$ nm, $y=0.2$ QWs demonstrating
increasing spin precession frequency with higher Mn doping and an
enhanced transverse spin lifetime with light Mn doping, which
decreases with higher doping. }\end{figure}

Measurements of $N_0 \alpha$ are made using results from both
time-resolved Kerr rotation (KR) and SIMS as described in Myers
\textit{et al}.\cite{Myers:2005} and Poggio \textit{et al}.
\cite{Poggio:2005}.  Pulses from a mode-locked Ti:sapphire laser
with 76-MHz repetition rate are split into pump and probe beams with
average powers of 2 mW and 0.1 mW respectively; the beams are
focused to an overlapping spot on the sample with the laser
propagation direction perpendicular to the external magnetic field
$B$ ($x$ axis) and parallel to the QW growth axis ($z$ axis).
Changes in the linear polarization angle (measured as KR) of the
reflected probe beam are measured as a function of the time delay
between the two pulses, producing a signal proportional to the
pump-induced electron spin polarization.  Fits to a decaying cosine
function yield the transverse spin coherence time $T^{*}_{2}$ and
the electron Larmor precession frequency $\nu_L$ (Fig. 2b).  As
observed in Ref. \onlinecite{Poggio:2005}, light Mn doping increases
$T^{*}_{2}$ within each ($d$, $y$) set.  Using the methods of Refs.
\onlinecite{Myers:2005} and \onlinecite{Poggio:2005}, the $\nu_L$
are converted to spin splittings $\Delta E$ which are fit to $\Delta
E = g_e \mu_B B - x N_0 \alpha \langle S_x \rangle$, where $g_e$ is
the in-plane electron g-factor, $\mu_B$ the Bohr magneton, and
$\langle S_x \rangle$ is the spin of the paramagnetic Mn system
along the applied field.  In this way a single value for $N_0
\alpha$ is extracted for each ($d$, $y$) sample set from the KR
(Fig. 3).

\begin{figure}\includegraphics{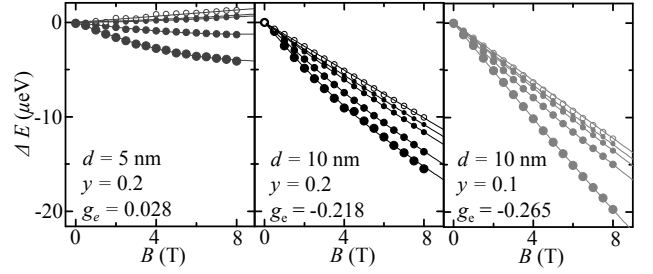}\caption{\label{fig3}
Spin splittings extracted from KR for $d=10$ nm, $y=0.1$ (gray),
$d=10$ nm, $y=0.2$ (black), and $d=5$ nm $y=0.2$ (light gray) with
increased dot size representing higher Mn concentration.
Non-magnetic control samples (open symbols) give the electron
$\textit{g}$-factor used in the spin-$\frac{5}{2}$ paramagnetism
fits (solid lines).\cite{Myers:2005} }\end{figure}

Plots of $N_0 \alpha$ as a function of $y$ for $d= 5$ nm and $d= 10$
nm (Fig. 4, inset) show that for constant $d$ a larger barrier
height leads to a more negative exchange constant.  We plot $N_0
\alpha$ as a function of $E_{e}$ in Fig. 4.  The $s$-$d$ exchange
becomes more antiferromagnetic with increasing $E_e$; this holds
equally well for variation in $E_e$ due to changes in both $d$ and
$y$.  Note especially that samples with the same $d$ (same symbol)
but different $y$ (different $E_e$) are consistent with the $E_e$
dependence.

\begin{figure}[b]\includegraphics{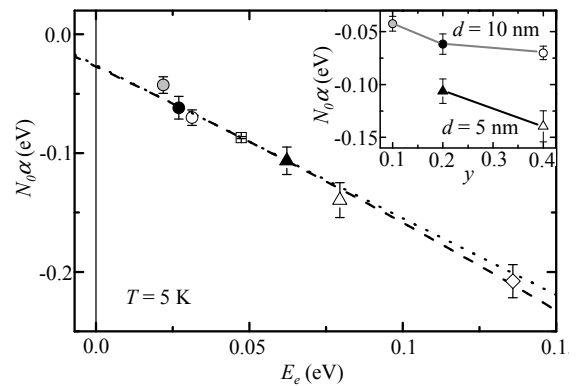}\caption{\label{fig4}
$N_0 \alpha$ plotted as a function of electron kinetic energy for
QWs with $d = 10$ nm (circles), $d = 7.5$ nm (squares), $d = 5$ nm
(triangles), and $d = 3$ nm (diamonds).   $y = 0.4$ (open symbols)
are from Ref. \onlinecite{Myers:2005}, while $y = 0.2$ (black) and
$y = 0.1$ (gray) are new to this manuscript.  The dotted line is the
linear approximation and the dashed line is an envelope function
calculation based on Ref. \onlinecite{Merkulov:1999}. The inset
compares $N_0 \alpha$ for QWs of different barrier height $y$ but
same width.  }\end{figure}

Interactions between dilute Mn spins and carrier spins are typically
treated with a Kondo exchange Hamiltonian $H_{sp-d} = - \sum_{i}
J_{sp\text{-}d} \sigma \cdot S_i$ where $\sigma$ is the carrier
spin, $S_i$ is the spin of a Mn moment, and $J_{sp\text{-}d}$ is the
exchange constant.\cite{Larson:1988}  Direct Coulomb exchange
contributes a positive (ferromagnetic) interaction to
$J_{sp\text{-}d}$.  Kinetic exchange due to virtual transitions
between band states and localized $d$ states causes an interaction
which is equivalent (by a Schrieffer-Wolff transformation) to a
negative (antiferromagnetic) exchange contribution to
$J_{sp\text{-}d}$.\cite{Schrieffer:1966, Blinowski:1992} These
virtual transitions, depicted in Fig. 5, typically dominate in the
valence band where $d$ levels strongly hybridize with the band-edge
hole states, leading to antiferromagnetic coupling between the hole
and Mn ion spins.  In the conduction band, kinetic exchange vanishes
at $\textbf{k} = 0$ by symmetry,\cite{Dietl:1994} leaving
ferromagnetic direct exchange  dominant.

This band-edge picture, supported by numerous magneto-optical
measurements in bulk II-VI DMS, is inadequate for describing the
reduced dimensionality of exchange in quantum-confined
heterostructures.  One approach to this situation is to treat the
exchange constant $J_{sp\text{-}d}$ as $\textbf{k}$-dependent within
$\textit{k}\cdot\textit{p}$ theory, which leads to a reduction in
the magnitude of valence band kinetic exchange with increasing
quantum confinement energy.\cite{Bhattacharjee:1998}  Though this
model replicates the qualitative features of the data in II-VI
materials, the prediction for the reduction of $N_0 \alpha$ is a
factor of $\sim 5$ smaller than what is observed in CdMnTe QWs.
\cite{Mackh:1996, Bhattacharjee:1998}   In addition, this model
amounts to a reduction of $|N_0 \alpha|$ with increasing $k$, which
is inconsistent with our findings in GaMnAs QWs.

The negative shift in the antiferromagnetic $N_0 \alpha$ observed in
our experiments is better explained by a more complete model of the
effects of reduced dimensionality accounting for the admixture of
$p$-symmetry valence band states and $s$-symmetry conduction band
states for non-zero ${\textbf k}$ arising in
$\textit{k}\cdot\textit{p}$ theory.\cite{Merkulov:1999}  As quantum
confinement increases electron kinetic energy (and thus $k$), the
electron wavefunction takes on more $p$-symmetry character,
increasing the antiferromagnetic contribution to $N_0 \alpha$ from
kinetic exchange.  As Merkulov \textit{et al.} point
out\cite{Merkulov:1999}, depending on the degree of the admixture, a
positive conduction band exchange constant can decrease and even
become negative with increasing kinetic energy.

We quantitatively apply the envelope function model of Ref.
\onlinecite{Merkulov:1999} to the data.  To lowest order in $E_e$,
the dimensionless slope of $N_0 \alpha$ vs. $E_e$ is given by:
\begin{eqnarray}
\label{eq1} \frac{d (N_0 \alpha (E_e) ) }{ d |E_e|}  = - \frac{ 2
(E_{\text{g}}+ \Delta)^2 + E^{2}_{\text{g}} }{E_{\text{g}}
(E_{\text{g}} + \Delta) (3 E_{\text{g}} + 2 \Delta) } \times  \{ N_0
\alpha_{\text{pot}} \nonumber\\ - [N_0 \beta_{\text{pot}} + N_0
\beta_{\text{kin}} \gamma ] \times [1 - \frac{4 \Delta^2}{3[2
(E_{\text{g}} + \Delta)^2 + E^{2}_{\text{g}}]} ] \}
\end{eqnarray}

where $\Delta$ is the spin-orbit coupling, $E_g$ is the band gap,
$N_0 \alpha_{\text{pot}}$ ($N_0 \beta_{\text{pot}}$) is the direct
conduction (valence) band edge exchange integral, and $N_0
\beta_{\text{kin}}$ is the kinetic exchange integral of the valence
band.  The conduction band kinetic exchange integral is assumed to
be zero in the vicinity of $k = 0$. $\gamma$ is a parameter (see
Ref. \onlinecite{Merkulov:1999}) that corrects the
$\beta_{\text{kin}}$ given by Schrieffer-Wolff kinetic exchange in
the valence band for electrons in the conduction band.  This
correction depends on the energies for virtual hole and electron
capture in the Mn $d$ levels, $\epsilon_{+}$ and $\epsilon_{-}$,
respectively (Fig. 5).  The energy levels for electron and hole
capture in the Mn $d$ levels are estimated from the
configuration-interaction analysis of core-level photoemission
spectroscopy where the core ground state is assumed to be the
Mn$^{2+}$ ion (Fig. 5a).\cite{Okabayashi:1998}  The $d^5$ core and
valence band hole $\underline{L_0}$ hybridize with the
$d^6\underline{L_0}^2$ state and the $d^4$
state,\cite{Mizokawa:1997} leading to virtual transition energies in
a one-particle picture of $\epsilon_{+} = -5.2 \pm 0.5$ eV and
$\epsilon_{-} = 2.7 \pm 0.5$ eV relative to the valence band edge
(Fig. 5b).  $\gamma (E_e)$ accounts for the energy shift to the
conduction band shown in Fig. 5c.   For the model, valence band
potential exchange is assumed insignificant ($N_0 \beta_{\text{pot}}
= 0$) while valence band kinetic exchange is estimated as $ N_0
\beta_{\text{kin}} = -1.2$ eV based on photoemission and transport
measurements\cite{Okabayashi:1998, Omiya:2000} and from
calculations.\cite{Bhattacharjee:2000b} $N_0 \alpha_{\text{pot}}$ is
the only free parameter in the model, appearing both in Eq. 1 and as
the y-intercept value of $N_0 \alpha (E_e)$ at $E_e = 0$.  Eq. 1 is
generally insensitive to $N_0 \alpha_{\text{pot}}$ since $N_0
\beta_{\text{kin}}$ is typically an order of magnitude larger, so
the extrapolation to $E_e = 0$ determines the fit value for $N_0
\alpha_{\text{pot}}$.

\begin{figure}\includegraphics{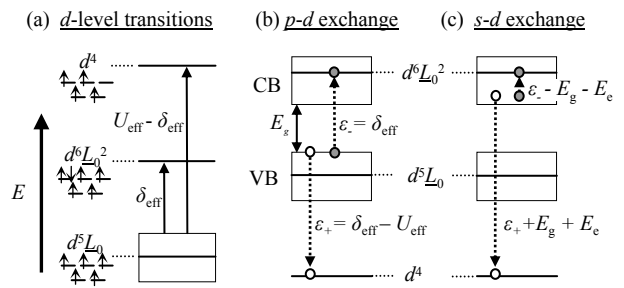}\caption{\label{fig5}
(a) Schematic of the configuration-interaction model for core-level
transitions from a $d^5$ Mn core with a valence band hole
($d^5\underline{L_0}$) to a $d^6$ core with two valence holes
($d^6\underline{L_{0}^2}$) or a $d^4$ core.  The effective
parameters $U_{\text{eff}} - \delta_{\text{eff}}$ and
$\delta_{\text{eff}}$ measure energies from the bottom of the core
multiplet to the valence band edge.\cite{Okabayashi:1998} (b) The
same core-level scheme shifted into a one-particle picture.  The
transitions between Mn core levels correspond to the electron and
hole capture energies, $\epsilon_{-}$ and $\epsilon_{+}$,
responsible for valence band kinetic exchange.  (c) The
corresponding scheme adjusted for kinetic exchange involving
conduction band electrons; $\gamma (E_e)$ in Eq. 1 corrects for the
energy difference between the carriers in each scheme.
}\end{figure}

Both the linear approximation of Eq. 1 and a more detailed
calculation of the envelope function theory of Ref.
\onlinecite{Merkulov:1999} are shown in Fig. 4. The slope calculated
from Eq. 1 is $-1.3 \pm 0.3$, agreeing well with a best-fit slope of
$-1.42 \pm 0.12$.  The model extrapolates to a bulk value of $N_0
\alpha$ = $-25 \pm 1$ meV.  The envelope function theory calculation
accounts for the weak $E_e$ dependence in $\gamma (E_e )$, yielding
curvature at high $ E_e$.  Because $E_e$ is small compared to
$\epsilon_{-}$ and $\epsilon_{+}$, the curvature is too small to be
directly observed by our measurements.  Attempts to observe
curvature at high $E_e$ would be further complicated by the
increased wavefunction penetration into the AlGaAs barriers.  Since
the carrier virtual capture energies are not well known in AlGaAs,
reliable estimates of the effects of barrier penetration are not
possible and calculations within this envelope theory are not
adequate at high $E_e$.  The good agreement between the model and
the data support the conclusion that band mixing dominates the
behavior $s$-$d$ exchange as a function of 1D confinement energy in
GaMnAs.  While other treatments of kinetic exchange in QWs, such as
the $sp^{3}$ tight binding model,\cite{Bhattacharjee:2000} are not
excluded by this experiment, these models in their current form do
not, to our knowledge, provide such numerical agreement in both
II-VI and III-V DMS experiments.

Though accounting for the confinement effect, the above discussion
does not address the extrapolation of our measurements to a bulk
antiferromagnetic GaMnAs exchange of $N_0 \alpha = -25 \pm 1$ meV.
This is an order of magnitude smaller than that typically measured
in II-VI materials, and of the opposite sign.  Its magnitude is
similar to the +23 meV measured by Raman spin-flip scattering in
GaMnAs,\cite{Heimbrodt:2001} but the sign is inconsistent with both
previous measurements and band-edge $s$-$d$ exchange theories.  We
must therefore consider that some of the assumptions which are valid
in II-VI do not apply in GaMnAs. For instance, because the Mn$^{2+}$
ions replace Ga$^{3+}$ in the lattice, they represent a repulsive
Coulomb potential for the conduction band electrons.
\cite{Sliwa:2005}  The resulting charge screening alters the overlap
of Mn and carrier wavefunctions and is so far neglected in
traditional calculations of $sp$-$d$ exchange.   The density of
defects should be highly sensitive to growth conditions, leading to
large variation in the charge screening and hence the measured
values of the exchange constants.

In summary, we have extended earlier investigations into the kinetic
energy dependence of $s$-$d$ exchange in GaMnAs/AlGaAs QWs to
include variations of both the QW width and barrier height. This
relationship is found to be well fit by a band-mixing model where
confinement-induced $p$-symmetry in the conduction band causes an
antiferromagnetic contribution to $s$-$d$ exchange.  The barrier
height dependence of the kinetic exchange mechanism demonstrates
that the exchange interactions in III-V magnetic heterostructures
can be tuned using low-dimensional band-engineering of quantum
confinement.

\begin{acknowledgments}
We thank T. Dietl and J. A. Gaj for enlightening discussions and J.
H. English and A. W. Jackson for MBE technical assistance.  This
work was supported by DARPA, ONR, and NSF. N.P.S. acknowledges
support from the Fannie and John Hertz Foundation.
\end{acknowledgments}


\begin{thebibliography}{26}

\bibitem{Dietl:1994} T. Dietl, (Diluted) Magnetic Semiconductors, in Handbook of Semiconductors, ed. S. Mahajan, Vol.3B (North-Holland, Amsterdam, 1994) p. 1251.
\bibitem{Awschalom:1985} D. D. Awschalom, J.-M. Halbout, S. von Molnar, T. Siegrist, and F. Holtzberg, Phys. Rev. Lett. \textbf{55}, 1128 (1985)
\bibitem{Crooker:1997} S. A. Crooker, D. D. Awschalom, J. J. Baumberg, F. Flack, and N. Samarth Phys. Rev. B \textbf{56}, 7574 (1997); S. A. Crooker, J. J. Baumberg, F. Flack, N. Samarth, D. D. Awschalom, Phys. Rev. Lett. \textbf{77}, 2814 (1996).
\bibitem{Ohno:1996} H. Ohno, A. Shen, F. Matsukura, A. Oiwa, A. Endo, S. Katsumoto, and Y. Iye, Appl. Phys. Lett. \textbf{69}, 363 (1996); H. Ohno, D. Chiba, F. Matsukura, T. Omiya, E. Abe, T. Dietl, Y.Ohno, and K. Ohtani, Nature \textbf{408}, 944 (2000).
\bibitem{Myers:2005a} R. C. Myers, K. C. Liu, X. Li, N. Samarth, and D. D. Awschalom, Phys. Rev. B \textbf{72}, 041302R (2005).
\bibitem{Gaj:1979} J. A. Gaj, R. Planel, G. Fishman, Solid State Commun. \textbf{29}, 435 (1979).
\bibitem{Twardowski:1983} A. Twardowski, T. Dietl, M. Demianiuk, Solid State Commun. \textbf{48}, 845 (1983).
\bibitem{Larson:1988} B. E. Larson, K. C. Hass, H. Ehrenreich, and A. E. Carlsson, Phys. Rev. B \textbf{37}, 4137 (1988).
\bibitem{Blinowski:1992} J. Blinowski and P. Kacman, Phys. Rev. B \textbf{46}, 12298 (1992).
\bibitem{Mackh:1996} G. Mackh, W. Ossau, A. Waag, and G. Landwehr, Phys. Rev. B \textbf{54}, 5227R (1996).
\bibitem{Bhattacharjee:1998} A. K. Bhattacharjee, Phys. Rev. B \textbf{58}, 15660 (1998).
\bibitem{Merkulov:1999} I. A. Merkulov, D. R. Yakovlev, A. Keller, W. Ossau, J. Heurts, A. Waag, G. Landwehr, G. Karczewski, T. Wojtowicz, and J. Kossut, Phys. Rev. Lett. \textbf{83}, 1431 (1999).
\bibitem{Myers:2005} R. C. Myers, M. Poggio, N. P. Stern, A. C. Gossard, and D. D. Awschalom, Phys. Rev. Lett. \textbf{95}, 017204 (2005).
\bibitem{Poggio:2005} M. Poggio, R. C. Myers, N. P. Stern, A. C. Gossard, and D. D. Awschalom, Phys. Rev. B \textbf{72}, 235313 (2005).
\bibitem{Szczytko:1996} J. Szczytko, W. Mac, A. Stachow, A. Twardowski, P. Becla, and J. Tworzydlo, Solid State Commun. \textbf{99}, 927 (1996).
\bibitem{Ando:1998} K. Ando, T. Hayashi, M. Tanaka, and A. Twardowski, Appl. Phys. Lett. \textbf{83}, 6548 (1998)
\bibitem{1D:Poisson} One-dimensional Poisson-Schr\"{o}dinger solver written by G. Snider (http://www.nd.edu/gsnider/)
\bibitem{Plot:1986} B. Plot, B. Deveaud, B. Labert, A. Chomette, and A. Regreny, J. Phys. C \textbf{19}, 4279 (1986).
\bibitem{Schrieffer:1966} J. R. Schrieffer and P. A. Wolff, Phys. Rev. \textbf{149}, 491 (1966).
\bibitem{Bhattacharjee:2000} A. K. Bhattacharjee and J. P$\acute{\text{e}}$rez-Conde, \textit{Proc. of 25th Int'l. Conf. Phys. Semicond.}, eds. N. Miura and T. Ando (Springer-Verlag, Berlin, 2000) p. 242.
\bibitem{Okabayashi:1998} J. Okabayashi, A. Kimura, O. Rader, T. Mizokawa, A. Fujimori, T. Hayashi, and M. Tanaka, Phys. Rev. B \textbf{58}, R4211 (1998).
\bibitem{Mizokawa:1997} T. Mizokawa and A. Fujimori, Phys. Rev. B \textbf{56}, 6669 (1997).
\bibitem{Omiya:2000} T. Omiya, F. Matsukura, T. Dietl, Y. Ohno, T. Sakon, M. Motokawa, and H. Ohno, Physica E \textbf{7} 976 (2000).
\bibitem{Bhattacharjee:2000b} A. K. Bhattacharjee and C. Benoit \`{a} la Guillaume, Solid State Commun.  \textbf{113}, 17 (2000).
\bibitem{Heimbrodt:2001} W. Heimbrodt, Th. Hartmann, P. J. Klar, M. Lampalzer, W. Stolz, K. Volz, A. Schaper, W. Treutmann, H.-A. Krug von Nidda, A. Loidl, T. Ruf, V. F. Sapega, Physica E \textbf{10}, 175 (2001).
\bibitem{Sliwa:2005} C. $\acute{\text{S}}$liwa and T. Dietl, preprint, arXiv:cond-mat/0505126

\end{thebibliography}
\end{document}